\documentstyle[prb,aps,prl,multicol]{revtex}
\begin{document}
\draft
\tighten   
\title{A Transfer Matrix Study of the Staggered BCSOS Model}
\author{Enrico Carlon \cite{newaddress}}
\address{Instituut voor Theoretische Fysica, Universiteit Utrecht,
         Postbus 80006, 3508 TA Utrecht, Nederland\\
	 and
	 HLRZ, Forschungszentrum J\"ulich, D-52425 J\"ulich, Deutschland}
\author{Giorgio Mazzeo}
\address{Dipartimento di Fisica, Universit\`a di Genova,
 	 and Istituto Nazionale per la Fisica della Materia (INFM),\\
         via Dodecaneso 33, 16146 Genova, Italia}
\author{Henk van Beijeren}
\address{Instituut voor Theoretische Fysica, Universiteit Utrecht,
         Postbus 80006, 3508 TA Utrecht, Nederland\\
	 and
	 HLRZ, Forschungszentrum J\"ulich, D-52425 J\"ulich, Deutschland}
\date{\today}
\maketitle
\begin{abstract}
The phase diagram of the staggered six vertex, or body centered solid
on solid model, is investigated by transfer matrix and finite size
scaling techniques.
The phase diagram contains a critical region, bounded by a 
Kosterlitz-Thouless line, and a second order line describing a deconstruction
transition. In part of the phase diagram the deconstruction line and
the Kosterlitz-Thouless line approach each other 
without merging, while the deconstruction changes its critical behaviour
from Ising-like to a different universality class. Our model has the
same type of symmetries as some other two-dimensional models, such as
the fully frustrated XY model, and may be important for understanding
their phase behaviour.
The thermal behaviour for weak staggering is intricate. It may be relevant
for the description of surfaces of ionic crystals of CsCl structure.
\end{abstract}
\pacs{PACS numbers: 68.35.Rh, 64.60.Cn, 64.60.Fr, 68.35.Bs}

\begin{multicols}{2}
\section{INTRODUCTION}
\label{sec:intro}

Six vertex models were introduced by Slater \cite{Slter} to describe 
ferroelectricity in two dimensional networks.  Placing arrows on the
bonds of a square lattice one can define the sixteen possible
arrangements of arrows pointing towards and away from a lattice point as
vertices. In six vertex models only those six vertex configurations
are kept (see Fig.\ \ref{FIG01}) that satisfy the {\em ice rule}, i.e.
they have two arrows pointing in and two pointing out at each vertex.
Assigning energies $\epsilon_1$, \ldots, $\epsilon_6$ to these vertices
one obtains a class of exactly solved models \cite{LiebWubook,Baxter}.

Six vertex models can also be interpreted as surface models, by mapping them
to the so-called body centered solid on solid (BCSOS) models
\cite{HvBPRL}, defined as limiting cases of
a lattice gas, or Ising model, on a body centered cubic lattice. Therefore
the phase structure of the six vertex model as function of its vertex
weights can be translated directly to the surface phase structure of
the corresponding BCSOS model. 
The mapping turned out to be important in understanding the properties
of the {\em roughening transition} \cite{HenkIm}. Using the exact 
solution of the six vertex model it was found that roughening is a 
transition of infinite order of Kosterlitz-Thouless (KT) type, 
confirming previous renormalisation group results \cite{ChuiWeeks}.

Experimental situations often are too complex to allow even a qualitative
description by the exactly solved BCSOS models. Various extensions of the 
standard six vertex model have been proposed to deal with these cases. Two
main classes may be identified: one where interactions 
between vertices are added, and another one in which the vertex lattice
is split into two sublattices with different vertex energies.
These modifications, however, lead to models which, apart from 
some particular cases \cite{BaxFreeFer,FreeFer}, lose the property
of being exactly solvable. Other techniques (e.g. numerical ones) have
to be adopted.
Models in the first class have been proposed to account for further 
neighbour interactions between surface atoms, which may change 
the symmetry of the ground state and give rise to phase 
transitions other than the roughening transition. Vertex interactions were 
introduced to reproduce the $(2 \times 1)$
reconstruction of the (110) face of fcc noble metals like Au and Pt 
\cite{LeviTouz}. This led to
investigations on equilibrium phase transitions on these surfaces as well
as on surfaces of lighter metals like Ag, Rh, etc... \cite{MJLT}.
A model of the (100) surface of an fcc crystal exhibiting a 
$(2 \times 2)$ reconstructed ground state \cite{BastKnops} has 
recently extended the list.
The second class of models, with vertex weights alternating on the two 
sublattices, are known as {\em staggered six vertex models}. 
A staggering only involving the weights of vertices 5 and 6 corresponds
to the imposition
of a ``staggered field'', i.e. a field coupled to the arrow directions that
changes sign between neighbouring arrows. This gives rise to an inverse 
roughening transition in part of the phase diagram \cite{Erik}.
Alternating the values for the energies of the vertices 1, 2 and 3, 4 on the
two sublattices leads to a model known as ``the staggered six vertex model''
(or staggered BCSOS model) in the literature. In a large part of its
parameter space it can be mapped onto the Ashkin-Teller model \cite{atdual}.
Using this transformation Knops investigated the phase structure of the 
staggered BCSOS model in part of its phase diagram by
renormalisation group methods \cite{Knops}, but until recently a large
region of the phase diagram has remained unexplored.

In this paper we present a complete account of our investigations,
over the full range of parameters, of the staggered BCSOS model. A preliminary 
description has been given already in Ref.\ \cite{ourPRL},
here we present further details as well as a number of new results. 

In the unexplored region of the phase diagram the model has a ground
state which is twofold degenerate, therefore
it has a symmetry of Ising type. The twofold degeneracy is lost
at a second order transition line which approaches another line of
KT roughening transitions. The interplay between the two is particularly
interesting, especially since a similar interplay between a KT and a
second order transition
has been found for several different models, among which other models for
reconstructed surfaces \cite{MJLT,denNijs3}, but also the fully 
frustrated XY model \cite{ffXY1,ffXY2,Olsson} and coupled XY -- Ising models
\cite{XYising1,XYising2}.
They have received a great deal of attention in recent years and till now 
their critical behaviour is not fully understood. The strong
interplay between Ising and KT degrees of freedom may lead to several
possible scenarios where, in a certain region of the phase
diagram, either the two transitions occur close to each other 
but remain separate, or they merge into a single phase transition,
which may perhaps belong to a new universality class. 

Apart from these more theoretical aspects the model is likely to be
relevant for the study of the equilibrium properties of a certain 
class of crystal surfaces, e.g. the (001) surface of ionic crystals
of the CsCl structure. This too will be discussed in some detail. 

The paper is structured as follows. In Section II we give a description
of the model. In Section III we present its full phase diagram. 
In Section IV we review the techniques employed in our studies, i.e. the 
transfer matrix
method and finite size scaling, and discuss the correlation functions and
free energies we
calculated to derive our results. 
In Section V we discuss the critical exponents of the model and some
possible scenarios for the changes in the critical behaviour along the
deconstruction line. In Section VI we conclude with a brief discussion
of related models.

\section{The staggered six vertex model}

The partition function of the six vertex model is given by
\begin{eqnarray}
Z=\sum_{\{\cal C\}} \,e^{\textstyle{- \beta \sum_{i=1}^6 
n_i({\cal C}) \, \epsilon_i}}  
\label{partfun}
\end{eqnarray}
where the sum runs over the set of all allowed vertex configurations 
$\{{\cal C} \}$ and $n_i({\cal C})$ denotes the number of vertices
of type $i$ in the configuration $\cal C$
($\beta = 1 /k_B T $, with $k_B$ 
Boltzmann's constant and $T$ 
the temperature). 
The model has been solved exactly \cite{LiebWu} for any 
choice of values of the energies $\epsilon_i$ ($i=1 \ldots 6$).
A relatively simple choice of the vertex energies is given by
$\epsilon_1 = \epsilon_2 = \epsilon_3 = \epsilon_4 = \epsilon$
and $\epsilon_5 = \epsilon_6 = 0$ which defines, for $\epsilon
>0$, the so-called F model. The ground state is twofold degenerate
and is composed of vertices 5 and 6
arranged alternatingly in a chess board configuration. The low temperature
phase is usually called an ``antiferroelectric" phase, since along both
horizontal and vertical rows the arrows predominantly alternate in direction.
From the exact solution it is known that this system undergoes an
infinite order phase transition to a disordered paraelectric state
at $\beta \epsilon = \ln 2$.

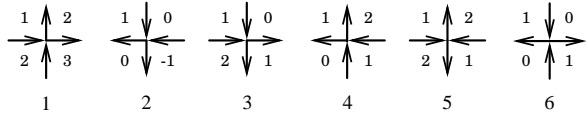
\begin{figure}[h]
\caption{The six vertices and the corresponding height
configurations.}
\label{FIG01}
\end{figure}

As already pointed out in the introduction, the six vertex models are
isomorphic to a class of solid on solid (SOS) models called BCSOS models
\cite{HvBPRL}. Microscopic configurations of an SOS model are given
in terms of discrete heights $h_i$ of surface atoms with respect to a
reference plane. All lattice sites up to these heights are occupied and
all sites above them are empty.
In the BCSOS model the height variables are placed on the dual lattice
of the six vertex lattice.
This is subdivided into an even and an odd sublattice, which are
intertwined in a chess board pattern and on which the surface heights
assume even respectively odd values only. The even sites will be referred
to as black (B) sites and the odd ones as white (W) sites.
In addition the height differences between neighbouring sites are
restricted to the values $\pm 1$.  
The mapping of six vertex configurations to corresponding configurations
of a BCSOS model is very simple.
The height differences between neighbouring sites are put in a one-to-one
correspondence with the arrow directions in the six vertex configuration.
The convention is that the
higher of the two surface sites is at the right side of the arrow.
Given a configuration of vertices, the configuration of heights
is fixed uniquely once the height of a reference atom has been
fixed (see Fig.\ \ref{FIG01}).
 
The vertex energies can be reinterpreted in terms of bond energies between
the atoms.
When periodic boundary conditions are applied along the (say)
horizontal direction of the vertex lattice the number of
vertices 5 and 6 per row is equal, therefore with no loss of
generality one can always choose $\epsilon_5 = \epsilon_6 = 0$,
fixing the point of zero energy.
The vertices 5 and 6 describe local configurations 
in which the height variables on either diagonal are equal 
(see Fig.\ \ref{FIG01}).
Vertices 1, 2, 3 and 4 correspond to configurations where 
the height variables along either of the two diagonals are 
different, therefore
$\epsilon_1$, $\epsilon_2$, $\epsilon_3$ and $\epsilon_4$ can be
viewed as energies needed to break a next nearest neighbour bond and
produce a height difference of two vertical lattice units between
neighbouring sites of equal colour.

In the ordinary
BCSOS model the distinction between B and W atoms has been 
introduced only for convenience of description, but the two 
sublattices are equivalent and are treated exactly on the
same footing. 
Knops \cite{Knops} extended the model to a two component system
where the B and W atoms are physically different.
While energy zero is still attributed to all vertices 5 and 6,
Knops assigned two different energies, $\epsilon$ and 
$\epsilon^{\prime}$, to broken bonds between W-W and B-B atoms
respectively. In terms of the six vertex representation also the
vertex lattice is divided into two alternating sublattices
I and II 
on which the vertices assume different energies as follows:
\end{multicols}
\begin{eqnarray}
\left\{\begin{array}{cccr}
& \mbox{on sublattice I:} \; &\epsilon_1=\epsilon_2=\epsilon;\;\; 
                             \epsilon_3=\epsilon_4=\epsilon^{\prime};\;\; 
                             \epsilon_5=\epsilon_6=0 &
\\
& \mbox{on sublattice II:} \; &\epsilon_1=\epsilon_2=\epsilon^{\prime};\;\;
                             \epsilon_3=\epsilon_4=\epsilon;\;\;
                             \epsilon_5=\epsilon_6=0 & \;.
\end{array}\right.
\label{energies}
\end{eqnarray}
This choice defines the staggered six vertex model. In the BCSOS
representation the model is described by the hamiltonian:
\begin{eqnarray}
H &=& \frac{\epsilon}{2} \sum_{\langle ij \rangle}
                                \left\vert h_i^W - h_j^W \right\vert +
  \frac{\epsilon^\prime}{2}\sum_{\langle kl \rangle} 
                                \left\vert h_k^B - h_l^B \right\vert
\label{hamiltonian}
\end{eqnarray}
\begin{multicols}{2}
subject to the constraint that the height difference between
neighbouring B and W sites is $\pm 1$. The first sum in 
(\ref{hamiltonian}) runs over all pairs of neighbouring
W sites 
on the surface
and the second sum over the corresponding B pairs. Throughout 
this article we will also use the parameter $\delta$, defined
by the relation $\epsilon^{\prime}=\epsilon + 2 \delta$.
As mentioned already in the Introduction the model defined here will
be referred to as the staggered six vertex (or BCSOS) model. Obviously when 
$\epsilon = \epsilon^{\prime}$ ($\delta =0$) one recovers the usual
F model.

For negative values of the vertex energies $\epsilon$ and
$\epsilon^{\prime}$, the system may model ionic crystals of bcc 
structure as for instance CsCl \cite{simple}. The constraint of minimal
height difference between neighbouring surface sites reflects the
effects of the strong attraction between oppositely charged ions, while
neighbouring pairs of the same colour, having
equal charges, repel each other. It is further assumed that on top
of the Coulombic repulsion other interactions, as for instance spin
exchange, generate a slight difference in the energies for broken
bonds between B-B and W-W pairs ($\epsilon \neq \epsilon^{\prime}$).
In the staggered BCSOS model the interaction range is limited to next
nearest neighbours and to have a more realistic representation of
ionic crystals one needs to extend the interactions
to further neighbours. Yet we expect the phase structure described here
for the staggered BCSOS model may be encountered in real ionic crystals.

\section{THE PHASE DIAGRAM}
\label{sec:phase}

We have investigated the phase diagram of our model by means of
transfer matrix and finite size scaling techniques, which will be
the subject of Section \ref{sec:numerical}.
Here we present the main results. Since
the model shows a trivial symmetry upon exchange of $\epsilon$ and
$\epsilon^{\prime}$, corresponding to the replacement $(\delta,\epsilon)
\longleftrightarrow (-\delta,\epsilon+2\delta)$, we can
restrict ourselves to the region $\delta \ge 0$. 
The phase diagram naturally divides into three sectors of globally different 
behaviour, though smoothly connected to each other. These are described in
the three following subsections.

\subsection{The range {\mbox{
\boldmath
{$ \; \epsilon > 0; \; \epsilon^\prime > 0 $}}}}

For positive values of $\epsilon$ and $\epsilon^{\prime}$ Knops
\cite{Knops} investigated the phase diagram through a mapping onto the 
Ashkin-Teller model \cite{atdual}; the phase diagram of the
latter had been obtained before by renormalisation group methods \cite{ATRG}.
On the $\beta \epsilon$ axis the (001) surface in the corresponding
BCSOS model is in a flat phase for $\beta \epsilon > \ln2$, whereas the
interval $\beta\epsilon \le \ln2$ represents the temperature region in
which the surface is rough. The infinite order transition occurring at
the KT point $\beta \epsilon = \ln 2$, $\beta \delta = 0$ corresponds
to the roughening transition of this surface.

Roughening is a phase transition which can be characterized by the 
vanishing of the free energy of a {\em step}, separating two surface
regions of different average height. The roughening transition results
into a proliferation of steps leading to a delocalisation of the
surface position 
and to a logarithmic divergence of the mean square height difference
at large distances:
\end{multicols}
\begin{eqnarray}
G(R_{ij})=\langle \left( h_i - h_j \right)^2  \rangle \; \sim 2 a_V^2 K(T) 
           \ln R_{ij}
           \;\;\;\;\mbox{for}\;\;\;\; \mbox{$R_{ij} \rightarrow \infty$}
           \;\;\;\;\mbox{and}\;\;\;\; \mbox{$T \ge T_R$}
\label{gofr}
\end{eqnarray}
\begin{multicols}{2}
with $a_V$ the vertical lattice spacing, $K(T)$ a temperature dependent
prefactor, $R_{ij}$ the distance between the lattice
sites $i$ and $j$ and $T_R$ the roughening temperature. Below $T_R$,
$G(R)$ saturates for large $R$ at a temperature dependent constant value.
Renormalisation group calculations \cite{Ohta} show that at $T_R$ the
prefactor assumes the universal value 
\begin{eqnarray}
K(T_R)=\frac{1}{\pi^2} \, .
\label{koft}
\end{eqnarray}
In the particular case of the exactly solved F model, $K(T)$ is known
for every temperature above $T_R$
\cite{axe}, that is
\begin{eqnarray}
K(T)=\frac{1}{\pi\arccos\Delta}
\label{univjump}
\end{eqnarray}
where $\Delta = 1 - e^{2 \beta \epsilon}/2$. In fact Eqs.\ (\ref{gofr}) and
(\ref{univjump}) are valid not only in the high temperature phase of the
F model, but also for
$\delta =0$, \ $\beta \epsilon < 0$, which defines the 
so-called inverted F model \cite{Erik}: all along the negative
$\beta \epsilon$ axis the surface is in a rough state.

For $\delta \neq 0$ Knops found two critical lines originating from
the KT point  and running into the regions $\delta > 0$ and $\delta < 0$.
The lines represent phase transitions of
Ising type from an ordered low temperature phase to a disordered
flat (DOF) phase, similar to the phase introduced by Rommelse and
Den Nijs \cite{DOF}.

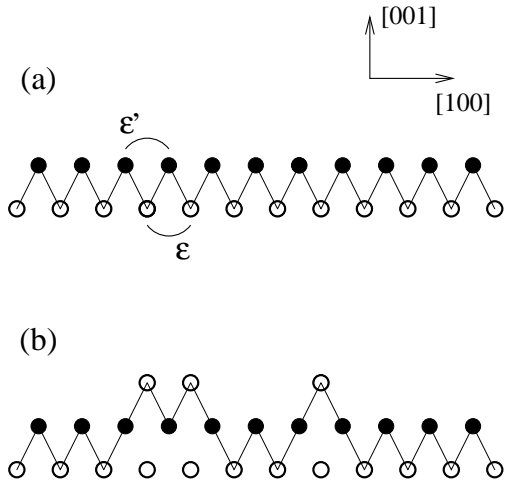
\begin{figure}[h]
\caption{Side view of the surface for $\epsilon > 0$ in one of its 
         ground states (a) and in the Ising limit $\beta \delta 
	 \rightarrow \infty$ at a finite value of temperature (b). 
         In the ground state atoms form uninterrupted rows also in the
         $[010]$ direction (orthogonal to the page). The structure 
	 of the DOF phase resembles that of (b), with broken bonds 
	 between B atoms always less frequent than between W atoms 
	 (their energies are $\epsilon^{\prime}$, $\epsilon$ 
	 respectively, with $\epsilon^{\prime} > \epsilon > 0$).
	 In the figure, $\epsilon$ and $\epsilon^{\prime}$ denote 
	 bonds the breakage of which would cost these amounts of 
	 energy.}
\label{FIG02}
\end{figure}

The ground state of the model is twofold degenerate. At higher
temperatures the more weakly bound sublattice fluctuates
above and below the more strongly bound sublattice, which remains
almost localised at a given level.
In the limit $\delta \rightarrow \infty$ the model can be mapped 
{\em exactly} onto the two dimensional Ising model, which is critical
at $\beta \epsilon = \ln(1 + \sqrt 2)$; the strong
sublattice is ``frozen" to height (say) zero, the only freedom left
for the heights of the other sublattice is to take the values $\pm 1$
just below or above that of the strong sublattice
(see Fig.\ \ref{FIG02}). 
According to the
renormalisation group results obtained for the Ashkin-Teller model
\cite{ATRG} the phase transition remains in the universality class
of the two dimensional Ising model all along the critical line down to
$\delta = 0$.

Starting from the low temperature phase and increasing the
temperature, the system 
undergoes an Ising transition to the DOF phase, while roughening
is pushed up to infinite temperature \cite{Knops}.
We reinvestigated this part of the phase diagram with the transfer
matrix methods to be described in Section \ref{sec:numerical} and
obtained results in full agreement with those of Knops.

\subsection{The range {\mbox{
\boldmath
{$ \; \epsilon < 0; \; 
\epsilon^{\prime} > 0 $}}}}

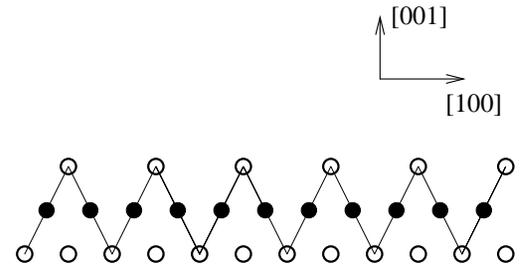
\begin{figure}[h]
\caption{Side view of one of the ground states of the model
         for $\epsilon < 0$. Notice that W atoms alternate in height
	 with respect to the B sublattice also in the $[010]$ direction
 	 (orthogonal to the page).}
\label{FIG03}
\end{figure}

When $\epsilon$ becomes negative the ground state of the system changes
drastically (irrespective of the sign of
$\epsilon^{\prime}$). Breaking bonds between white atoms now lowers the energy,
so that at zero temperature one finds the black sublattice unbroken
(provided $\delta >0$) while atoms of the white sublattice are found
alternatingly above and below the black sublattice (see 
Fig.\ \ref{FIG03}).
This surface configuration is commonly referred to as a $c(2\times 2)$
reconstructed surface. In the equivalent six vertex representation the
ground state is formed by columns of vertical arrows running alternately
all upwards and all downwards, and by rows of horizontal arrows running
alternately all right and all left.
Such an arrangement of directed paths is known as a {\em Manhattan
lattice}, due to its resemblance to the one-way street pattern 
of Manhattan.
As the energy is invariant under the reversal of all arrows, the 
ground state is twofold degenerate, just as in the case $\epsilon>0$. 
Indeed
in the limit $\delta \to \infty$ the model can be mapped {\em exactly}
onto an antiferromagnetic Ising model leading to the value
$\beta\epsilon=-\ln(1+\sqrt{2})$ for the critical temperature. This
constitutes a horizontal asymptote, as in the case $\epsilon > 0$,
for a second order transition line, whose existence can be deduced
again from the mapping of the staggered six vertex model onto the
Ashkin-Teller model \cite{critfan}.
It separates a low temperature $c(2\times 2$) reconstructed
phase from a high temperature DOF phase, where the reconstruction order 
is lost but the surface is still globally flat: it is the same DOF
phase found for $\epsilon > 0$; no singularities are met in crossing
the $\beta \delta $-axis.
Our transfer matrix calculations confirm the existence of this critical line and
show it exhibits Ising type critical behaviour
throughout sector B.

\subsection{The range {\mbox{
\boldmath
{$ \; \epsilon < 0; \; \epsilon^{\prime} < 0 $}}}}
\label{sec:subC}

For $\epsilon^{\prime} < 0 $ the mapping of the staggered six vertex
model to the Ashkin-Teller model leads to negative Boltzmann weights
in the latter.
It loses its physical relevance and cannot be used any more to make
predictions on the phase behaviour of the staggered six vertex model.
In spite of this Kohmoto {\em et al.} \cite{critfan} have made some
conjectures, which have proven to be correct, on the physical situation
beyond the ``horizon" $\epsilon + 2 \delta =0$.

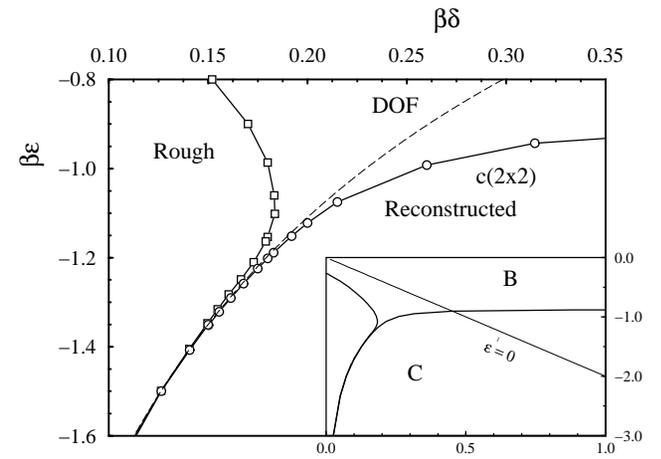
\begin{figure}[h]
\caption{The phase diagram of the staggered BCSOS model. We show 
	 here only the sector C, and part of the sector B in the 
	 inset. Open circles denote the deconstruction line and 
	 open squares the  roughening line. The estimate for the 
	 roughening condition, provided by Eq.\ (9), is shown as 
	 a dashed line. Is is almost indistinguishable from the 
	 correct curve (squares) for 
	 $\beta \epsilon \protect \lesssim -1.3$.}
\label{FIG04}
\end{figure}

Our transfer matrix analysis shows the existence of three phases: a low
temperature $c(2\times 2)$ reconstructed phase and a DOF phase, which are 
present already in sector B, and a rough
phase, which is found only in the present sector. Two critical lines
separate these phases, as shown in Fig.\ \ref{FIG04}: the first one is just
the continuation of the second order line beyond the horizon.
It still separates the  $c(2\times 2)$ region from the
DOF region and asymptotically approaches the axis $\beta\delta =0$. We
have strong indications that, within at least a major part of the sector
$\epsilon^{\prime} < 0 $, this line does not belong to the Ising
universality 
class. We will present the evidence for this in 
Section \ref{sec:exponents}.
The other critical line is a line of KT points separating the rough
region (or {\em critical fan}, as predicted already by Kohmoto {\em et al.}
\cite{critfan}) from the DOF region.

The point where the KT line meets the vertical axis can be determined
from the exact solution of the F model as the point where the prefactor
of the logarithmic term in the mean square height difference is
four times as large as its universal value assumed at the ordinary
roughening temperature of the F model
\begin{eqnarray}
K(T) &=& 4 K(T_R) = \frac{4}{\pi^2}
\label{factfour}
\end{eqnarray}
from which one obtains, inverting (\ref{univjump}):
\begin{eqnarray}
\beta \epsilon &=& \frac{1}{2} \ln\left(2 - \sqrt{2} \right)
\approx -0.2674 \, .
\end{eqnarray}
The factor four in (\ref{factfour}) stems from the fact that for
$\delta \neq 0$ the roughening transition is driven by steps of a
height of two vertical lattice units (as in Ref.\ \cite{Erik}), due
to the inequivalence between the two atomic sublattices.

    A simple estimate of the roughening transition temperature based on
    a random walk approximation (see Ref.\ \cite{ourPRL}) yields

\begin{eqnarray}
e^{-2\beta\delta}\,+\,e^{\beta\epsilon} &=& 1 \, .
\label{pottsline}
\end{eqnarray}
This line has been drawn in Fig.\ \ref{FIG04}. Indeed, for large and
negative $\beta\epsilon$ it is seen to run very close to
the KT line, which we could determine with great accuracy by the methods 
described in the next section.

A most remarkable feature of our phase diagram is the apparent 
merging of the second order and the KT line into a single line 
(see Fig.\ \ref{FIG04}).

Their horizontal distance $d$ as a function of $\beta\epsilon$ 
can be well described by a curve of the form
$d(\beta \epsilon) = C e^{\alpha \left( \beta \epsilon \right)} $,
with $\alpha \simeq 12$ \cite{ourPRL}.
This exponential fit suggests that although
the two lines are coming rapidly closer together as $\beta \epsilon$
is decreasing, they do never actually merge. Other fits, of the form
$d(\beta\epsilon) = C \left\vert \beta\epsilon - \beta\epsilon_0
\right\vert ^ {\alpha}$, which would be expected to work in the case
of a merging of the lines at $\epsilon=\epsilon_0$, could not be
stabilised against changes in the fitting range.

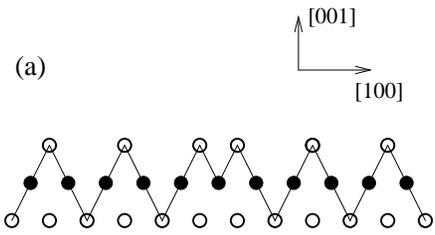
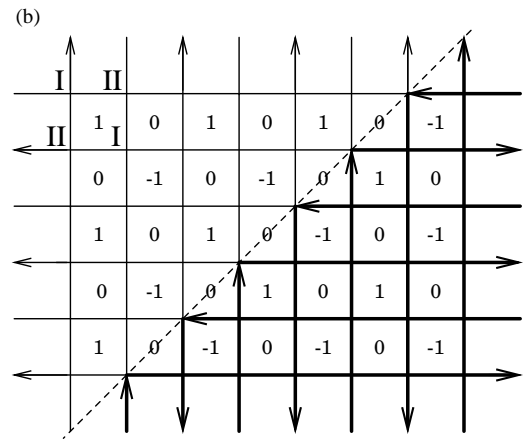
\begin{figure}[h]
\caption{A domain wall separating two different Ising phases 
	 for $\epsilon < 0$, (a) in side
         view and (b) seen from above in the vertex lattice.}
\label{FIG05}
\end{figure}

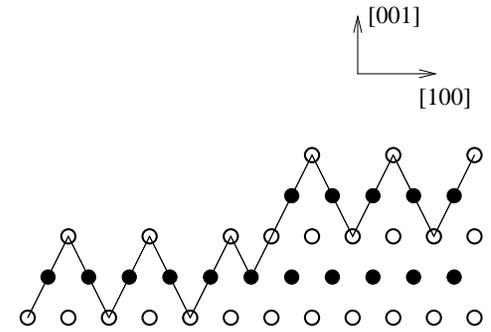
\begin{figure}[h]
\caption{Side view of a (double height) step as an excitation
	 of the Manhattan ground state.}
\label{FIG06}
\end{figure}

The apparent non-crossing of the two critical lines at first looks
very surprising. At low temperatures a domain wall between two
different Ising phases mainly consists of diagonal sequences of
vertices 5 and 6, as depicted in Fig.\ \ref{FIG05}; its energy
per unit length approximately equals $-\epsilon/\sqrt{2}$.
On the other hand a step consists mainly of long horizontal and
vertical chains of overturned arrows 
and has an approximate energy
per unit length of $2\delta$.
To a first approximation steps do not couple with the Ising order,
since the reconstructed phase remains the same at both sides of the
step 
(see Fig.\ \ref{FIG06}).
Hence one would expect the KT line (characterized by vanishing step
free energy) and the Ising line (vanishing Ising domain wall free
energy) to cross near $\epsilon+2\sqrt{2} \, \delta=0$. We think that
the actual non-crossing of the two lines can be explained as follows. 
When temperature is raised, more and more closed steps will be formed
on the surface as one approaches the roughening temperature 
$T_R$.
On these steps the direction of the arrows is reversed. 
In this way the Ising order parameter becomes more and more diluted,
which will, eventually, strongly reduce the free energy of a domain
wall. If in the end the closed steps become so prolific that they
cover on average half of the surface, without becoming of infinite
length, the Ising order disappears without roughening of the surface.

For $2\delta \lesssim 0.4|\epsilon|$ the thermal behaviour implied
by our phase diagram is quite intricate and remarkable. At low
temperatures the surface is in a $c(2 \times 2)$ reconstructed flat
phase, then on raising the temperature there is a second order
transition
to a DOF phase, rapidly followed by a KT transition to a rough phase.
Next there is a reentrant KT transition to the DOF phase.
This is an inverted roughening transition similar to the one described
in \cite{Erik}. Finally, as temperature approaches infinity, the
system asymptotically approaches a rough phase again.
Instead for $2\delta \gtrsim 0.4 \vert \epsilon \vert$ the system goes
through a single phase transition from the ordered to a DOF phase and
remains flat for all finite temperatures.

\section{TRANSFER MATRIX AND FINITE SIZE SCALING METHODS}
\label{sec:numerical}

Transfer matrix techniques are frequently used in studies of the critical
properties of two dimensional systems with short range interactions.
The construction of the transfer matrix (TM) follows a standard
procedure and the interested reader is referred to the
existing literature \cite{Baxter} for details. 

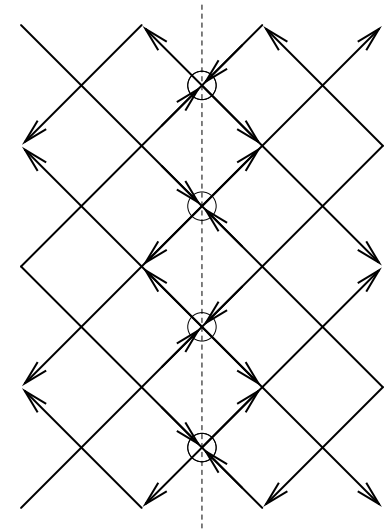
\begin{figure}[h]
\caption{Part of a ground state configuration of an $N \times 
	 \infty$ system with $N$ odd. Due to partial frustration, 
	 the system produces a domain wall made of a sequence of 
	 vertices 5 and 6 (denoted by circles).}
\label{FIG07}
\end{figure}

We use two different transfer matrices, one oriented parallel to 
the axes of the vertex lattice and another one tilted of $45^\circ$ 
with respect to these axes. We refer to the former as {\em vertical}
TM and to the latter as {\em diagonal} TM (Fig.\ \ref{FIG07} shows 
a configuration of the diagonal TM). 
We consider a lattice of width $N$ and height
$M$, with periodic boundary conditions in the horizontal direction.
For the vertical TM the subdivision of the lattice into a white and a black
sublattice, combined with the periodic boundary conditions, restricts $N$ to
even values.
For the diagonal 
TM the horizontal and vertical axis are chosen
along the diagonals of the vertex lattice and $N$ can be odd as well as even.
The element $T_{ij}$ of the matrix is defined as the Boltzmann
weight of a row of $N$ vertices generated by arrow configurations
labeled by the indices $i$ and $j$. One has $T_{ij}=0$ if this
row of vertices does not satisfy the ice rule. For the vertical TM,
if $i$ and $j$ are identical there are in fact two possible configurations
of rows of vertices: in this case the transfer matrix simply sums their
Boltzmann weights.
 
There are $2^N$ different arrow configurations for the vertical TM,
whereas for the diagonal TM this number is $2^{2N}$. The largest values
of $N$  we could treat numerically were $N=22$ for the vertical and
$N=12$ for the diagonal TM. Actually, due to the rotation of the lattice
over $45^\circ$ the latter should be compared to $12 \sqrt{2} \approx 17$ for
the vertical TM.
 
In the limit $M \rightarrow \infty$ the partition function per 
row becomes:
\begin{eqnarray}
\lim_{M \rightarrow \infty} \left( Z_{N \times M} \right)^{1/M}
=\lambda_0(N)
\end{eqnarray}
with $\lambda_0(N)$  the largest eigenvalue of $T$, from which the free
energy per row follows as
\begin{eqnarray}
\beta f(N) = - \ln \lambda_0(N) \, .
\label{free}
\end{eqnarray}

To each state $i$ we associate a polarisation $P_i = N_{i \uparrow}
- N_{i \downarrow}$, with $N_{i\uparrow}$ and $N_{i \downarrow}$ the total
numbers of up and down \cite{note2} arrows in the state $i$.
By virtue of the periodic boundary conditions in the horizontal
direction the transfer matrix can be reduced to blocks of fixed
polarisation, since $T_{ij} = 0$ if $P_i \neq P_j$ (see, for
instance, Ref.\ \cite{Baxter}).
The so-called central block is the one corresponding to zero
polarisation and describes a flat surface.
The blocks with polarisation $P_i = \pm 2$ (subcentral blocks) describe
a surface with a step.

The difference between the free energies of a surface with a step and of
a flat surface gives the step free energy, which, per unit of length, on
an $N \times \infty$ strip can be expressed as:
\begin{eqnarray}
\beta f_S(N) &=& - \left( \ln \lambda_1(N) - \ln \lambda_0(N) \right)
\label{STEPF}
\end{eqnarray}
where $\lambda_{0}(N)$ and $\lambda_{1}(N)$ are the largest
eigenvalue of the central and the subcentral block respectively 
\cite{fnlambda}.
The study of this quantity will allow us to determine the roughening
temperature.

The deconstruction transition can be studied by considering two
correlation lengths, which are both defined within the central block.
We define the inverse correlation length $\xi_D^{-1}$ as:
\begin{eqnarray}
\xi_{D}^{-1}(N) &=& -(\ln \lambda_2(N) - \ln \lambda_0(N) )
\label{defcsiN}
\end{eqnarray}
where $\lambda_{2}(N)$ is the second largest eigenvalue
of the central block. 
The other correlation length can be calculated from the diagonal TM
as the inverse of the domain wall free energy 
per unit length $f_W^{-1}(N)$, where $f_W(N)$ is given by
\begin{eqnarray}
f_W(N) &=& f(N) - \frac{f(N+1) + f(N-1)}{2}
\label{deffw}
\end{eqnarray}
with $N$ odd. Indeed in the diagonal TM an $N \times \infty$ strip, 
with $N$ odd, is partially frustrated since it cannot accommodate the 
Manhattan ground state without creating a domain wall 
(see Fig.\ \ref{FIG07}).
$f_W^{-1}(N)$ can be interpreted as the correlation length connected to 
the correlation function between two disorder variables \cite{KadCeva}.

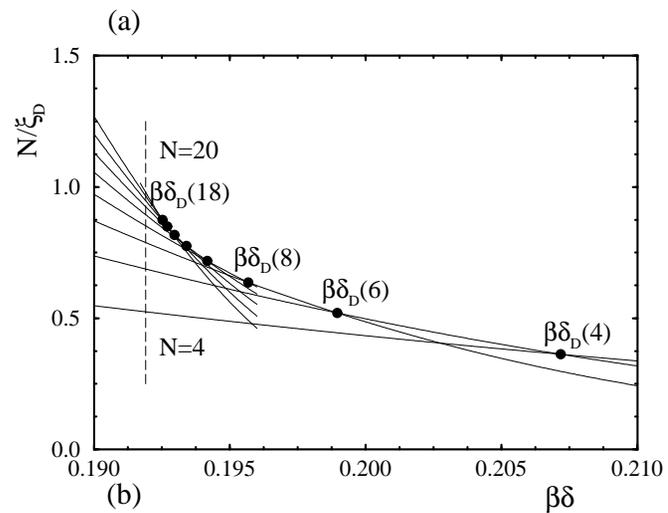
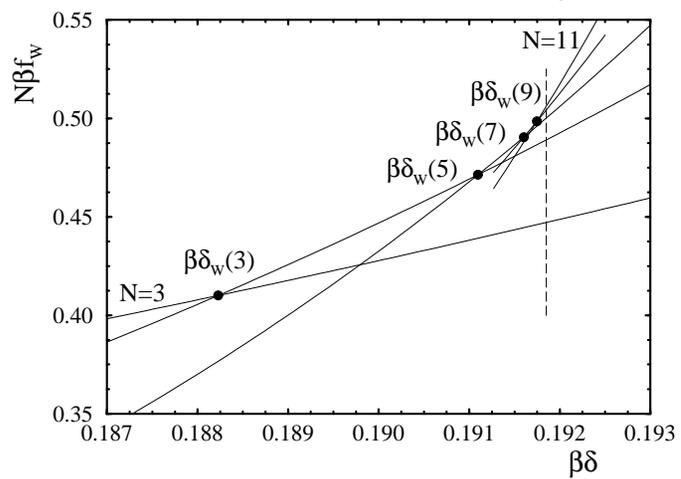
\begin{figure}[h]
\caption{The scaled correlation lengths $N \xi_D^{-1}(N)$ (a) 
	 and $N \beta f_W(N)$ (b) used to analyze the deconstruction 
         transition by means of the vertical respectively the 
	 diagonal transfer matrix. The curves for different system 
	 sizes intersect in a sequence of points which for 
	 increasing $N$ extrapolate to the deconstruction transition 
	 temperature.
	 The figures refer to a scan along the thermal trajectory 
	 characterized by $\epsilon/\delta=-6.0$.
	 The extrapolated values for (a) and (b) (dashed lines) 
	 coincide within error bars (not shown).}
\label{FIG08}
\end{figure}

Conformal invariance \cite{Cardybook} predicts that, at a critical point, 
the correlation lengths scale as $N$, so the deconstruction
transition can be located at the crossing point of the curves representing
the scaled quantities $N/\xi_D$ and $N \beta f_W$ as functions of the
temperature for different sizes.
In reality, as shown in 
Figs.\ \ref{FIG08}(a) and (b), no perfect crossing is found.
Instead, pairs of curves obtained for sizes $N$ and 
$N+2$ intersect
each other in a sequence of points, $(\beta \delta_D(N),\beta \epsilon_D(N))$,
respectively $(\beta \delta_W(N),\beta \epsilon_W(N)) $, which converges to
the infinite system critical point $(\beta \delta_D,\beta \epsilon_D)$,
respectively $(\beta \delta_W,\beta \epsilon_W)$.
An extrapolation procedure requiring several iterations \cite{iteratedfit}
is then used to estimate $\beta\delta_D$ or $\beta\delta_W$.
Of course the two independent estimates of the critical point have
to coincide, which provides a good check on the internal consistency
and accuracy of our procedures.

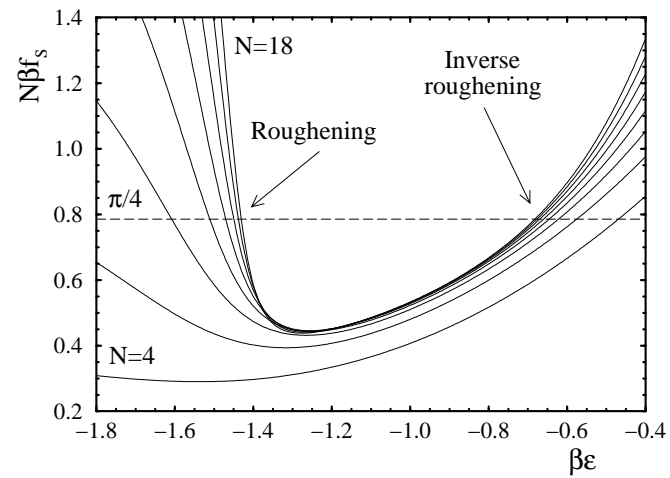
\begin{figure}[h]
\caption{The scaled step free energy $N \beta f_S(N)$ for different 
	 system sizes along a vertical scan across the phase 
	 diagram ($\beta \delta = 0.14$). Coalescence of curves is 
	 an indication that the surface is in a rough state. Both 
	 the roughening transition and the inverse roughening 
	 transition are visible: they 
	 can be roughly localized in the regions where the curves 
	 approach the value $\pi/4$.}
\label{FIG09}
\end{figure}

To locate the roughening temperature one has to employ a different method.
The scaling $f_S(N) \sim 1/N$ holds
not only at the KT transition
but also inside the rough region, where the surface is in a critical 
state. There the curves $N f_S(N)$, 
plotted as functions of temperature for
different values of N, coalesce in the limit $N \rightarrow \infty$ and
the point where they detach from each other can be identified as the KT
point (see Fig.\ \ref{FIG09}).
For an accurate location of $T_R$ one has to use the universal
properties of the KT transition which give rise to the scaling
prediction \cite{Erik,logcorr}

\begin{eqnarray}
N \beta f_S(N) = \frac{\pi}{4} + \frac{1}{A + B \ln N} 
\label{ktscaling}
\end{eqnarray}
which holds exactly at $T = T_R$,
with $A$ and $B$ 
nonuniversal constants. 
The constant $\pi/4$ is characteristic for steps with a height
of two vertical lattice spacings. The free energy of such a step
corresponds to the line tension between a vortex-antivortex pair
with vorticity 2 in the dual representation \cite{Erik}.
The KT transition temperature is
determined by requiring that a three-point fit of the form
$N \beta f_S(N) = A_0+1/\left(A+B\ln N\right)$ yields $A_0 = \pi/4$. For
the extrapolation we used iterated fits in the spirit of 
Ref.\ \cite{iteratedfit}.
We performed this procedure along different lines across the phase diagram,
scanning lines with $\beta \delta$ fixed, lines with $\beta \epsilon$ fixed
and thermal trajectories.

\section{CRITICAL EXPONENTS AND CENTRAL CHARGE}
\label{sec:exponents}

As we noted in the previous section, the critical line separating the
flat from the rough region can be well characterized as a KT line.
As we will see, the critical properties of the second order line are
less well determined, especially in the region $\epsilon^\prime < 0$.
We will calculate critical exponents and central charge pertaining to
the deconstruction transition using finite-size scaling methods.
The two exponents $\alpha$ and $\nu$ are related to the behaviour of
the singular part of the surface
free energy $f_{sing} \sim
t^{2-\alpha}$ and of the domain wall free energy $f_W \sim t^{\nu}$ 
\cite{othernu} (where $t = (T-T_D)/T_D$, $T_D$ the deconstruction
temperature). They satisfy the finite size scaling predictions
\begin{eqnarray}
\frac{1}{N}
\frac{\partial^2 f(N)}{\partial t^2} \sim N^{\frac{\alpha}{\nu}}
\label{specheat}
\end{eqnarray}
and
\begin{eqnarray}
N \frac{\partial f_W(N)}{\partial t} \sim N^{\frac{1}{\nu}}
\label{oneovernu}
\end{eqnarray}
respectively, valid at the critical point $T=T_D$. Two other critical indices we
will calculate are
\begin{eqnarray}
         x & = & \left. \frac{1}{2 \pi} \lim_{N\to\infty} \frac{N}{\xi_D(N)}
                                                   \right\vert_{T=T_D}
\label{xandx}
\\ \nonumber
\\
x^{\prime} & = & \left. \frac{1}{2 \pi} \lim_{N\to\infty} N
                                                       \beta f_W\left(N\right)
                                                   \right\vert_{T=T_D}
\label{xandxprime}
\end{eqnarray}
which represent the exponent of the spin-spin correlation function
\cite{eta} and that of the correlation function between disorder variables
\cite{KadCeva} respectively.
    The numerical errors on the values assumed by these quantities are obtained 
    as follows. We first evaluate the error on the determination of the
    critical temperature $\Delta T_D$ from the quality of the extrapolation
    to $N \to \infty$ of our finite size data \cite{iteratedfit}.
    Subsequently, we extract the values of the exponents,
    again by 
    iterated fits,
    at three different temperatures:
    $T_D - \Delta T_D$, $T_D$ and $T_D + \Delta T_D$.
    This procedure allows to 
    determine the maximum possible variation
    on the values of $\alpha/\nu$, $\nu$, $x$ and $x^\prime$,
    thus assigning them an error bar.
    Notice these errors are typically
    small if the critical temperature is determined accurately enough.

Finally, from conformal invariance \cite{Cardybook,BloCarAffl}
it follows that the leading finite size correction to the free
energy per site of an infinite system with periodic boundary
conditions, 
$\widetilde f_{\infty}$, is determined by the central charge (or
conformal anomaly) $c$ as
\begin{eqnarray}
\frac{f(N)}{N} \approx \widetilde f_{\infty} + \frac{\pi c}{6 N^2} \, .
\label{cc}
\end{eqnarray}
In fact we analyzed the central charge using the finite size approximation
 
\begin{eqnarray}
c(N,N+2) = \frac{3}{2 \pi}
           \frac{N^2 \left( N+2 \right)^2}{\left(N+1\right)}
           \left(\frac{f(N)}{N} - \frac{f(N+2)}{N+2}\right)
\label{ccc}
\end{eqnarray}
which converges to $c$ in the limit $N \to \infty$.

With the techniques described above we find
that the deconstruction line for $\epsilon < 0$ belongs no doubt to the
Ising universality class in sector B of our phase diagram.
Good convergence with increasing size is obtained
for the critical exponents as well as for the central charge, the values
of which are:
\begin{eqnarray}
\alpha  =  0 \,\,\,\,\,\,\,\, \nu  = 1 
\,\,\,\,\,\,\,\, x  =  x^{\prime}  =  \frac{1}{8} 
\,\,\,\,\,\,\,\,  c  =  \frac{1}{2} \; .
\label{Isingexp}
\end{eqnarray}
In the region $\epsilon^{\prime} < 0$ the situation is less clear. The
convergence of the data with increasing system size is worse, the values
of some of the critical exponents seem to vary along the critical line and
the central charge cannot be determined with any great accuracy. Yet
our results seem to clearly rule out the possibility that the critical line
remains in the Ising universality class. We present the results for the
various exponents and for the central charge below and then draw some more
general conclusions.

\subsection{{The exponent}{\mbox{
\boldmath
{$ \; x $}}}}
\label{sec:espx}

In part of sector C of the phase diagram we find difficulties in
convergence for the
quantities extracted from the correlation length $\xi_D(N)$. 
Fig.\ \ref{FIG10}(a) shows the behaviour of the 
exponent $x$
obtained from Eq.\ (\ref{xandx}), only along part
of the deconstruction line. The extrapolation procedure to infinite
size is in fact far from trivial close to the horizon
$\epsilon + 2 \delta = 0$, where we find non-monotonic
behaviour with increasing size for $N/\xi_D(N)$
and even for the sequence $\beta \delta_D(N)$.
In order to give an estimate of the exponent $x$ nonetheless,
we looked at the quantity $x(N) \equiv N/(2 \pi \widetilde \xi_D(N))$,
where $\widetilde \xi_D(N)$ is the correlation length evaluated now
at the intersection points $(\beta \delta_D(N),\beta \epsilon_D(N))$.

Fig.\ \ref{FIG10}(b) shows some plots of $x(N)$ vs. $N$
along the deconstruction line. 
The curves 1, 2 and 3 refer to critical points in sector B
located on the deconstruction line at $\beta \delta = 0.88$,
$\beta \delta = 0.60$ and $\beta \delta = 0.45$. They
show a good convergence to the Ising exponent $x =1/8$.
The other curves, (4-10), refer to the values $\beta \delta=
0.37$, $0.31$, $0.28$, $0.25$, $0.23$, $0.21$, $0.20$, $0.19$
in sector C of the phase diagram.
As the system
size increases the curves (4-7) show a reentrant behaviour towards
the value $x = 1/8$. At values of $\beta \delta \lesssim 0.20$ we find
monotonic convergence again as function of system size, but to values
which vary continuously as shown in Fig.\ \ref{FIG10}(a).
The behaviour of this set of curves suggests that 
along any thermal scan 
in sector C of the phase diagram the quantity $x(N)$ will show
an asymptotic decrease after a maximum. The position
of the maximum
gradually shifts to higher values of size until
it exceeds the largest value accessible to our calculations
and eventally disappears from sight.
As already mentioned, no accurate fit can be performed on 
curves 4, 5, 6 and 7 of Fig.\ \ref{FIG10}(a),
though a rough estimate provides values of $x$ below $1/8$.
When a fitted value can be extracted again (at smaller values
of $\beta \delta$) and drawn in Fig.\ \ref{FIG10}(a),
one should thus be cautioned against the possibility of
missing a maximum and a decreasing part.
This would provide values of $x$ possibly below $1/8$ and 
more in accordance with those of $x^{\prime}$ given in the following
sub-section.                                              
However, another difficulty may arise:
in the vicinity of the roughening transition
the correlation lenght $\xi_D$ may also be strongly influenced
by steplike excitations \cite{privdennijs}. A better quantity to look
at is represented by the exponent $x^{\prime}$.

\subsection{{The exponent}{\mbox{
\boldmath
{$ \; x^{\prime} $}}}}
\label{sec:espxp}

The quantity $N \beta f_W(N)$, converges monotonically as
function of the system size $N$ all along the deconstruction line.
For $0.25 \lesssim \beta \delta \lesssim 0.40$ the convergence is slow, but
it is still possible to give an estimate of the exponent $x^{\prime}$
using Eq.\ (\ref{xandxprime}). However the error bars are fairly large.
We notice a change in the direction of convergence: 
$N \beta f_W(N)/2 \pi$ converges to $x^{\prime}$ 
from above in sector B of the phase diagram
but from below in sector C.
Around the line $\epsilon^{\prime} = 0$ finite size effects are very small.
For $0.3 \lesssim \beta \delta \lesssim 0.4$ the exponent is still
compatible, within error bars, with the Ising value of $1/8$,
as shown in Fig.\ \ref{FIG11}, but for
$\beta \delta \lesssim 0.3$ the exponent shifts towards
values well below this.

\subsection{{The exponents}{\mbox{
\boldmath
{$ \; \alpha $}}}  
                      {and}{\mbox{
\boldmath
{$ \; \nu $}}}}

Fig.\ \ref{FIG12} shows the exponents $\alpha$ and $\nu$
calculated with the diagonal transfer matrix along the deconstruction
line, with the aid of standard extrapolation methods based on the
scaling relations (\ref{specheat},\ref{oneovernu}) \cite{Night}.
It is almost impossible to obtain these exponents using the vertical
transfer matrix, due to 
difficulties in convergence with increasing size.
These problems are much less severe with the diagonal
transfer matrix (see also the Appendix), even though the maximum available
system size is smaller.
The values thus obtained are not compatible with Ising
exponents when $\epsilon^{\prime} < 0$. They do satisfy
the hyperscaling relation $2 \nu = 2 - \alpha$ within error bars.

\subsection{The central charge}

In general the central charge $c$ vanishes in non-critical phases (here
the flat reconstructed phase and the DOF phase) and assumes finite values
at critical points or inside critical regions (like the rough phase).
As in the determination of the exponents
$\alpha$ and $\nu$, $c$ is calculated with the diagonal TM, as this leads
to better convergence and smaller finite size effects than calculations
with the vertical TM.
Fig.\ \ref{FIG13}(a) shows finite size approximations of $c$ along
vertical lines in the phase diagram based on Eq.\ (\ref{ccc}).
The left part of Fig.\ \ref{FIG13}(a) refers to a scan with $\beta
\delta =0.55$, which crosses the deconstruction line in a point 
of sector 
B, where we find exponents in the Ising
universality class. In this case the central charge at the
transition shows good convergence towards the Ising value ($c = 1/2$).
The right part of Fig.\ \ref{FIG13}(a) refers to a scan which
crosses the deconstruction line in a 
point of sector C with $\beta \delta = 0.25$.
Fig.\ \ref{FIG13}(b) shows two other plots of central charges
along vertical lines with $\beta \delta =0.22$ (left) and $\beta \delta
=0.20$ (right). In this part of the phase diagram the central charge
increases markably beyond the Ising value $c = 1/2$.
Due to strong finite size effects, slow convergence and nearness of the
KT line we cannot give a reasonable estimate for its actual value.

Fig.\ \ref{FIG13}(c) shows the central charge calculated along the
thermal trajectory $\epsilon + 10 \delta = 0$, starting from the rough
region (at small $\beta \delta$) towards the reconstructed phase at larger
$\beta \delta$.
According to our numerical results the line $\epsilon + 10 \delta = 0$
crosses the roughening and the deconstruction line in two points
very close in temperature.
In the infinite system limit the central charge should be 1 in the
rough region, drop abruptly from $1$ to $0$ at the KT point, remain $0$
in the DOF region, assume a non-zero value at the single point where 
the trajectory crosses the deconstruction line and remain $0$ beyond
that. In finite systems this behaviour is smeared out, as is the case
also in the other plots in Fig.\ \ref{FIG13}. Hence, since the KT
point and the deconstruction transition are extremely close on this
trajectory, one expects to see an apparent convergence of $c$ to the
sum of the KT value $1$ and that of the deconstruction transition.
For a deconstruction of Ising type this would yield $c=3/2$. From
conformal invariance \cite{Cardybook} it follows that for
unitary models with central charge smaller than unity $c$ can only assume
the values
\begin{eqnarray}
c &=& 1 - \frac{6}{M(M+1)} \;\;\;\; \mbox{with $M=3,4,\ldots \;\;$ .}
\label{discretec}
\end{eqnarray}
The Ising value $c=1/2$ is the lowest possible value, obtained with
$M=3$. Higher values of $M$ correspond to phase transitions in different
universality classes.
From Fig.\ \ref{FIG13}(c) it is apparent that $c$ converges to a
value larger than $3/2$, 
which we estimate around $c = 1.7$ --$1.8$.

\subsection{Deconstruction of non-Ising type?}

The results presented above strongly suggest that the
deconstruction transition is not in the Ising universality
class in, at least, part of the region $\epsilon^{\prime}<0$.
One cannot entirely exclude the possibility that the observed
deviations of critical exponents and central charge from their
Ising values are due to strong cross-over effects, induced by
the vicinity of the 
KT line \cite{privdennijs}, rather than being a genuine feature of
the deconstruction transition;
in view of our numerical results
however, we believe this is quite unlikely.

Of course, the next intriguing question is: what, if not Ising,
is the universality class of this reconstruction line? The answer
to this question is not easy and our numerical results are
not conclusive.

In general the exponents vary along the deconstruction line,
although some vary less than others. The exponent $x$ shows
generally worse convergence than the exponent $x^\prime$ and
extrapolation of the values of $x$ in part of the phase digram
turned out impossible due to the non-monotonic behaviour of the
finite size data as function of the system size $N$.

The exponent $x^\prime$ varies along the deconstruction line as
well, but it remains roughly constant in a limited region around
the value of $\beta \delta \approx 0.2$, with small error bars
thanks to rapid convergence of the finite size data.
At smaller values of $\beta \delta$ its value increases as well
as its error bars. This may be due to the vicinity of the KT line
or to the finite size effects caused by the increasing length of
straight step segments.
In general finite size effects increase at smaller values of
$\beta \delta$ (see also the Appendix); in this part of the
phase diagram the most important excitations consist of closed 
loops of reversed arrows which may become very elongated as
the energy per unit of length for a straight segment is
proportional to $2 \delta$.
One should expect that finite size effects are
particularly strong when the typical size of a loop becomes
of the same order of magnitude as the width of the 
strip, $N$.
Slow convergence also is present in a region to the left of the
line $\epsilon^\prime = 0$, as can be seen from the large
error bars around $\beta \delta \approx 0.3$ in Fig.\ \ref{FIG11}. 
This is due to a poor determination of the value of
the deconstruction temperature $T_D$.

The exponents $\alpha$ and $\nu$, as calculated from 
Eqs.\ (\ref{specheat},\ref{oneovernu}), 
vary along the deconstruction line in sector C. However,
the hyperscaling relation $2 \nu = 2 - \alpha$ is 
always satisfied within error bars.
In general, as shown in 
Fig.\ \ref{FIG12}, $\alpha$ tends to have
larger error bars than $\nu$.
In the region $\beta \delta \approx 0.2$, the convergence is rapid
in the sense that a two parameter fit is sufficient to extract
$\alpha$ and $\nu$ from (\ref{specheat}) and (\ref{oneovernu}).
At smaller values of $\beta \delta$ one in general needs to consider 
corrections to scaling using a three parameter fit.

Unfortunately our numerical results do not allow 
an inequivocal 
identification of the critical behaviour of the deconstruction
transition in the region $\epsilon^\prime < 0$. 
We notice however that the exponent $x^\prime$ remains constant in a
region around $\beta \delta 
\approx 0.2$, 
where the error bars are 
smallest.
In this region also $\alpha$ and $\nu$ converge rapidly
with increasing size, compared to other parts of the
deconstruction line in sector C.
One possible candidate for the observed exponents 
in this region could be that of the four
state Potts model, for which $\alpha = \nu = 2/3$, compatible
with our calculated values of $\alpha$ and $\nu$.

Conformally invariant models are classified according to the value
of their central charge, which can assume only discrete values
depending on some integer $M$, as given in Eq.\ (\ref{discretec}).
At fixed values of $M$ conformal invariance \cite{Cardybook}
predicts also the 
possible values for the exponents of 
correlation
functions at the critical point.
For the four state Potts model, the predicted exponents are
of the type 
$x, x^{\prime} = 2 p^2/q^2$ with $p$ and $q$ integers, as
pointed out in Ref.\ \cite{NienKnops}.
For $p=1$ and $q=4$ one indeed obtains the well-known magnetic
exponent $1/8$, instead for $p=1$, $q=5$ one obtains the value $2/25$.
Both values are shown as horizontal dashed lines in Fig.\ \ref{FIG11}; 
the exponent $x^{\prime}=2/25$ 
seems to fit
the measured values of the exponent very well for 
$\beta \delta \approx 0.2$. 
For the two dimensional Ising model conformal invariance predicts 
the exponents $x=1/8$ (magnetic) and $x=1$ (thermal) only.
Thus a measured exponent of 
value $x^{\prime} \approx 2/25$ 
is a quite clear sign of non-Ising critical
behaviour.
 
Further, the central charge clearly shifts away from its
Ising value $c=1/2$. For the four state Potts model we should
expect a central charge equal to $1$ ($M \to \infty$ in 
Eq.\ (\ref{discretec})).
The central charge markably increases in the region
$\epsilon^\prime < 0$. However, 
like for the critical exponents,
this increase goes smoothly from the Ising value, $c=1/2$, towards
higher values.
The central charge calculated along the line $\epsilon + 10
\delta = 0$, where the deconstruction and roughening line are
almost coinciding in temperature, 
extrapolates to $c = 1.7-1.8$, well above the Ising plus KT value
$c = 1/2 + 1 = 3/2$. As pointed out above,
this is 
another indication of non-Ising behaviour of the
deconstruction transition, 
though not quite compatible with that of
the four state Potts model, which would imply a central charge
equal to $c = 1 + 1 = 2$.

    Finally, also the possibility of having a line with continuously 
    varying exponents, as the behaviour of 
    especially the exponent $x$ in the
    sector C suggests (Fig.\ \ref{FIG10}(a)), should be considered.
    In this case the central charge would equal unity, as in the four 
    state Potts model.

    Anyhow, as discussed in Section V A - B, 
    the convergence of $x$ is much poorer than that of $x^{\prime}$.
    The slow shift of $x^{\prime}$ away from the Ising value as  
    $\beta \delta$ decreases in sector C (Fig.\ \ref{FIG11}) 
    is known to be a common feature of finite size scaling
    in the vicinity of points where a change of
    universality class 
    occurs.
    Moreover, as $\beta \delta$ becomes very small, the 
    nearness of the KT line is seen to influence the convergence of the 
    exponents of the deconstruction line. In conclusion, it seems
    quite plausible to have in practice only a window of 
    $\beta \delta$ values where constant critical exponents 
    are found. To enlarge this window one would have to consider larger system 
    sizes.

Bastiaansen and Knops \cite{BastKnops} recently studied a six vertex
model with 
an extended range of interactions.
They also found a phase diagram
with a second order line approaching a KT line. The exponents
of the second order line clearly deviate from their Ising
values and the authors 
suggested they might be explained as tricritical Ising
exponents. Applied to the staggered BCSOS model this
would mean a deconstruction line of Ising type with a
tricritical point, continuing beyond this point as a first
order line,
which is the phase behaviour of the annealed
diluted Ising (or Blume-Capel \cite{BlumeCapel}) model.
The exponents at the tricritical point would be
$\alpha = 8/9$, $\nu = 5/9$ and the central charge $c = 7/10$.
For the exponent $x^\prime $ conformal invariance predicts 
$x^\prime = 3/40$.
Around
$\beta \delta \approx 0.2$ the extrapolated value of $x^\prime$
would also be compatible with this value, but
$\alpha$ and $\nu$ are far away from their 
tricritical values.
At smaller values of $\beta \delta$ we do find exponents
which approach those of the tricritical Ising model, but
this happens in a region where the values we obtain for 
$x^\prime$ clearly shift away from  $3/40$ and where in
general finite size effects are quite strong. These same finite 
size effects also make it impossible to tell whether at
sufficiently small $\beta \delta$ the deconstruction line
becomes first order or not.

The point along the deconstruction line where the change of
universality class occurs is not sharply determined by our
numerical results.
We do not observe an abrupt jump of the exponents at a given
point, rather a continuous shift. A reasonable candidate for the
point separating the two regions (i.e. Ising and non-Ising),
could be the point where the deconstruction line crosses the line
$\epsilon^{\prime} = \epsilon + 2 \delta =0$. Crossing this line, we find
changes in the type of convergence of the exponents $x$ and 
$x^{\prime}$ (Sections V-A, B),
although without an abrupt change in their values.
We recall that in the surface representation of the model, in
one region the coupling constants between the atoms in the two
sublattices are both negative ($\epsilon < 0$, $\epsilon^{\prime}<0$);
in the other (where the
deconstruction transition is of Ising type) one of the two
coupling constants is positive ($\epsilon^{\prime}>0$).
In terms of the vertex lattice, in the region $\epsilon^{\prime}>0$,
vertices 5 and 6 are the excited vertices with the 
lower energy above the ground state 
value; at $\epsilon^{\prime}<0$, vertices
5 and 6 get the 
higher excitation energy.

All these considerations suggest that the properties of the
system may change between the two regions $\epsilon^\prime > 0$ and
$\epsilon^\prime < 0$ and make it more plausible that the shifts
in the exponents are not just due to cross-over, but also result
from a real change of universality class of the deconstruction
transition \cite{footnote}.

\section{DISCUSSION AND CONCLUSION}
\label{sec:discuss}

In this article
we studied the critical properties of the
staggered BCSOS model. Using transfer matrix techniques
we found two critical lines describing the deconstruction and
the roughening of the $(001)$ surface of a two component bcc
crystal.
 
The two lines approach each other in part of the phase diagram,
apparently without merging.
According to our results the deconstruction line in part of
the phase diagram changes its universality class from Ising to
non-Ising, although further investigations are needed to make
this point more convincing.
On the basis of the exponents we find, we conclude
that a possible universality class matching these exponents reasonably
well, in the region where the best convergence is found,
is that of the four state Potts model. 
Another possible scenario is that proposed by Bastiaansen and 
Knops \cite{BastKnops}. In their six vertex model with
interactions extended to further neighbours 
it is hard too to
distinguish between a single critical line and two lines
approaching each other, but remaining separate.
They find critical
exponents for the deconstruction transition clearly deviating
from the Ising values and conjecture that the observed
criticality could be explained as tricritical Ising
behaviour. 
The idea of a diluted Ising model is particularly attractive
in our case where, as we have seen, the deconstruction 
transition is, to all likeliness, the consequence of the 
dilution of the Ising order in the system caused by the 
formation of a large number of closed steps of finite length. 
Unfortunately we find little numerical evidence for this scenario. 
Finally, also the possibility of having a line with 
continuously varying exponents cannot be completely excluded.
Other models of reconstructed surfaces have been studied by several
authors. Den Nijs \cite{denNijs3} introduced a model 
that describes $(110)$ missing-row reconstructed surfaces of
some fcc metals ($Au$, $Pt$, \ldots). He found a deconstruction and
a roughening line merging into a single critical line, whith Ising
and KT behaviour simply superimposed. From his data, as presented
in the literature, it is not possible to really distinguish between
actual merging or mere rapid approach of the lines. A clear distinction
to our model is that in Den Nijs' model the deconstruction transition
remains of Ising type throughout.
Another class of models 
for the same metal surfaces has been developed
and extensively studied by the Trieste group \cite{MJLT,Santoro}.
Again a deconstruction line and a KT roughening line
are seen to approach each 
other. Depending on the microscopic details of the 
model,
the deconstruction
line keeps its Ising character either all along,
or up to a tricritical point where it changes to a first order line.

As mentioned in the Introduction there are several other two
dimensional models with KT and Ising degrees of freedom. One
that has received a lot of attention, starting from the beginning of the
last decade \cite{Teitel}, is the fully frustrated XY model, which 
describes certain two dimensional Josephson junction arrays.
The study of its critical behaviour has led to several different
conjectures about its universality class and critical exponents.
Several papers \cite{ffXY1,ffXY2} report non-Ising exponents and it
has been suggested that the model would belong to a novel type of
universality class.
To our knowledge, whether this type of universality class would
or would not coincide with that of some known models has not been
established yet.
In the most recent study concerning the fully frustrated XY model
Olsson \cite{Olsson} 
presents evidence of two separate
transitions, a KT and an Ising one where the former occurs at
somewhat lower temperature than the latter: $T_{KT} < T_{IS}$.
This would be in agreement with our results since the XY model can
be mapped 
onto a solid-on-solid model via a duality transformation
\cite{KnopsXY}, which maps the low temperature phase of one model
onto the high temperature phase of the other and 
vice versa.
Olsson's work suggests that the non-Ising exponents observed
by other groups are due to the failure of some finite-size scaling
hypothesis used in previous works.
In the staggered BCSOS model instead, we find clear evidence of
non-Ising exponents. Unfortunately there exists no exact mapping
between this model and
the fully frustrated XY model, 
therefore they 
may well be in different universality classes.
Yet, we hope that some of the ideas developed in
this paper to study the staggered BCSOS model, 
will be
generalised to other 
models so as to reach a deeper
understanding of their critical properties.

\acknowledgements

It is a pleasure to thank Paul Bastiaansen, Henk Bl\"ote,
Hubert Knops, Bernard Nienhuis and Marcel 
den Nijs for stimulating discussions.
Financial support permitting several meetings between the
authors of this paper is gratefully acknowledged.
In particular H.\ v.\ B. thanks the Centro di Fisica
delle Superfici e delle Basse Temperature (CNR), Genova, 
while G.\ M. thanks the Instituut voor Theoretische Fysica, Utrecht,
and acknowledges the kind hospitality of prof. Dietrich Wolf 
at HLRZ, J\"ulich.

\appendix
\section*{FINITE SIZE EFFECTS}	

We show here how the diagonal and the vertical transfer matrix
have different finite size effects in part of the phase diagram.
Consider first the phase point $\delta = 0$, $\epsilon
\rightarrow -\infty$.
Vertices 5 and 6 are absent and the partition function can be
calculated easily. Consider
lattices of size $N \times M$ with cylindrical geometry, that
is the $N$ vertices along a horizontal row are connected
to each other through periodic boundary conditions.
In the vertical transfer matrix there are in total $2^{N+M}$
configurations, and the free energy per site is given by:
\begin{eqnarray}
-\beta \widetilde f = \frac{N+M}{N M} \ln 2 
\label{finit1}
\end{eqnarray}
(the symbol $\widetilde f$ is used to distinguish the free
energy per site from $f$, the 
free energy per row).
In the thermodynamic limit $N$,$M \rightarrow \infty$ the free
energy per site vanishes.

However in transfer matrix calculations one takes the limit
$M \rightarrow \infty$ keeping $N$ finite; this gives a free
energy per site equal to:
\begin{eqnarray}
-\beta \widetilde f = \frac{1}{N} \ln 2 \, .
\end{eqnarray}
 
With the diagonal transfer matrix the total number of configurations
available is $2^{2N}$, since once the arrows on a row are fixed the
whole configuration is fixed. Repeating the same calculation as
done above one finds a free energy per site:
\begin{eqnarray}
-\beta \widetilde f = 0
\label{finit2}
\end{eqnarray}
independent of the value of $N$.
The conclusion is that the free energy shows finite size corrections
of the order $1/N$ in the vertical transfer matrix, while there are
no finite size effects for the diagonal transfer matrix.
In Fig.\ \ref{FIG14} we plot the free energy calculated along
the line $e^{\beta \epsilon} + e^{-2 \beta \delta} =1$ with the
vertical
and diagonal transfer matrices. The endpoint $e^{-2 \beta \delta} =1$
corresponds to the free energy which we calculated in
(\ref{finit1}) and (\ref{finit2}).
As can be seen from the figure, there is a wide area to the left
of this point where the free energies calculated from the vertical
transfer matrix show large finite size effects, while the convergence
is much faster for the diagonal transfer matrix. In both cases the
convergence is faster for $e^{-2 \beta \delta} < 1/2$, that is in
the region of the phase diagram where only a second order line is 
present. Obviously, for small values of $\delta$ the boundary 
effects are very strong, due to closed loops of reversed arrows 
winding around the cylinder,
i.e. the ribbon which constitutes our system with periodic
boundary conditions in the horizontal direction (see Fig.\ \ref{FIG15}).
These closed loops are more frequent in the vertical transfer matrix,
since one can reverse the arrows along a horizontal line with a cost
in energy of $2 \delta N$. Closed loops in the diagonal transfer
matrix require at least a vertex 5 and a vertex 6, and therefore
occur less frequently.

\end{multicols}

\begin{figure}[h]
\caption{(a) The exponent $x$ calculated along the deconstruction 
	 line, in the range of values of $\beta \delta$ where the
	 convergence is monotonic. (b) Non-monotonic behaviour of
	 $x(N)$ as a function of N: curves 1, 2 and 3
	 refer to critical points in sector B of the phase diagram (and 
	 tend to the value $1/2$), while all the other curves (4-10) 
	 refer to points in sector C (see text). For (4-10), due to the 
	 limitation in the maximum system size available
	 (larger than that shown here) it is almost impossible to 
	 obtain a good extrapolation for $N \rightarrow \infty$.}
\label{FIG10}
\end{figure}

\begin{figure}[h]
\caption{The exponent $x^{\prime}$ calculated along the 
	 deconstruction line. The horizontal dashed lines 
	 represent the Ising value 1/8 and the four state Potts 
	 exponent 2/25. Error bars smaller than the symbol size 
         are not shown.}
\label{FIG11}
\end{figure}

\begin{figure}[h]
\caption{The critical exponents $\alpha$ and $\nu$ calculated 
	 along the deconstruction line. The dashed line represents 
	 their value in the four state Potts model.}
\label{FIG12}
\end{figure}

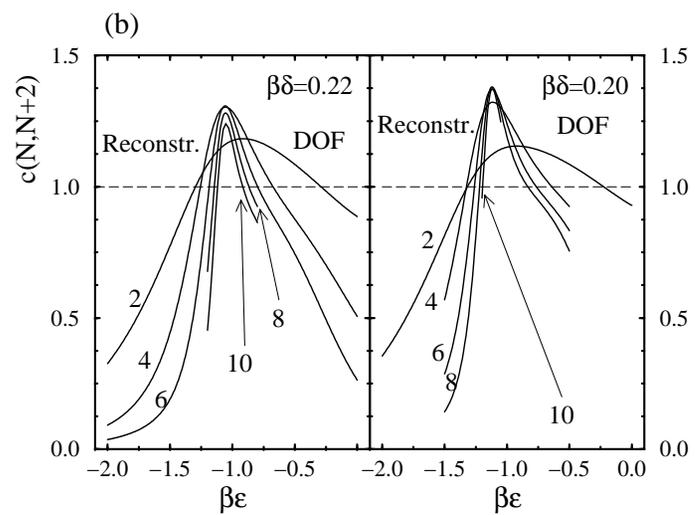
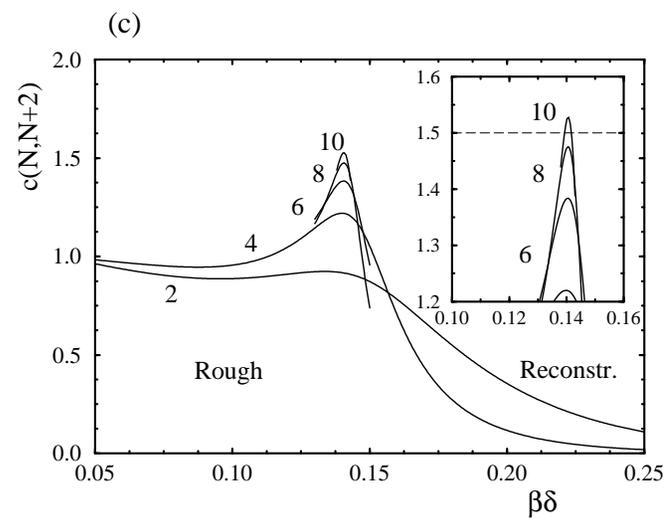
\begin{figure}[h]
\caption{Finite size approximations $c(N,N+2)$ of the central charge 
	 from Eq.\ (22) (numbers denote the system size $N$) across 
	 different phases in the phase diagram
	 calculated along vertical scans  at 
	 ((a), left) $\beta \delta =0.55$,
	 ((a), right) $\beta \delta =0.25$,
	 ((b), left) $\beta \delta =0.22$,
	 ((b), right) $\beta \delta =0.20$
	 and (c) along the thermal trajectory $\epsilon/\delta=-10.0$.
         The inset in (c) is just an enlargement of the same graph 
	 emphasizing the convergence of the central charge 
	 to values larger than 1.5.}
\label{FIG13}
\end{figure}

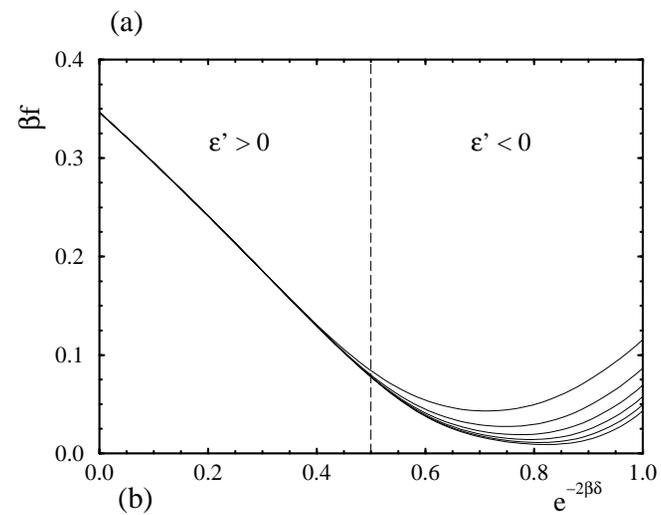
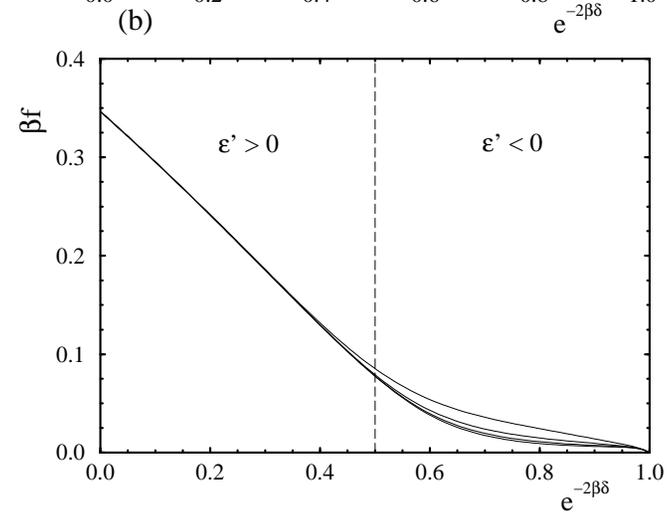
\begin{figure}[h]
\caption{Free energy along the line $e^{\beta \epsilon} + 
	 e^{-2\beta \delta} = 1$ (Eq.\ (9))
	 calculated (a) with the vertical transfer matrix and (b) 
	 with the diagonal transfer matrix for different system sizes.
         Notice the different scaling behaviour especially at the point 
         $e^{-2\beta \delta} = 1$: in (a) the curves 
         for increasing $N$ tend to the infinite system size value
         (zero) like $\ln 2/N$, in (b) they {\em are} zero. 
	 It is apparent that finite size 
	 effects are generally smaller in (b).}
\label{FIG14}
\end{figure}

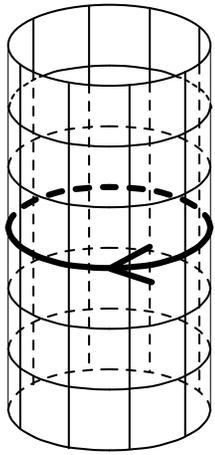
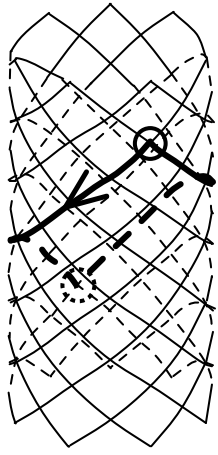
\begin{figure}[h]
\caption{When $\delta \ll \epsilon$, finite size effects may show 
	 up in the form of closed loops of reversed arrows winding 
	 around the cylinder. In the diagonal transfer matrix (b) 
	 however they are less frequent than in the vertical transfer 
	 matrix (a) since they require at least one couple of 
	 vertices 5 and 6 (circles) for closing up the lace.}
\label{FIG15}
\end{figure}


\begin{thebibliography}{99}
\bibitem{simple} Crystallographically $CsCl$ has a simple cubic structure;
however, if one ignores the identities of the different ions one obtains
a bcc lattice.
\bibitem{note1} The thermal trajectories are those trajectories 
                in the phase diagram obtained by fixing the
                values of the energies and varying the temperature.
\bibitem{eta} The exponent $x$ is related to the more standard
              exponent $\eta$ as $x=\eta/2$.
\bibitem{fnlambda} The largest eigenvalue of the transfer matrix
                   is always found in the central block.
\bibitem{footnote} Of course, if the second scenario, of a
diluted Ising line, would apply, the tricritical point is far
removed from  the point $\epsilon^{\prime}=0$ and there would have to
be strong cross-over effects.
\end{thebibliography}
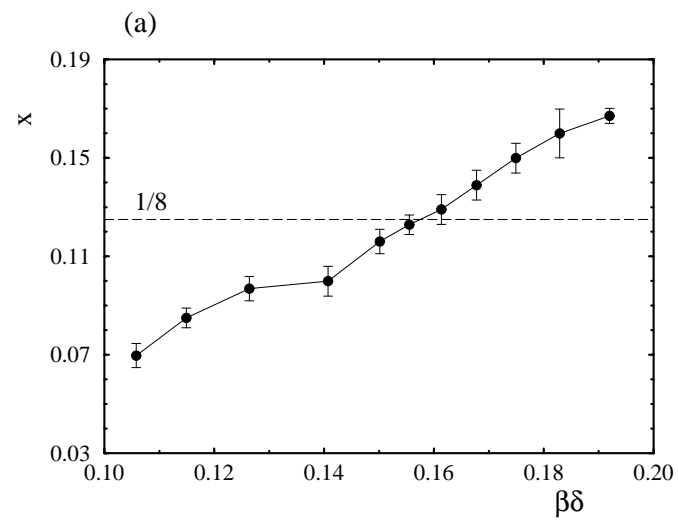
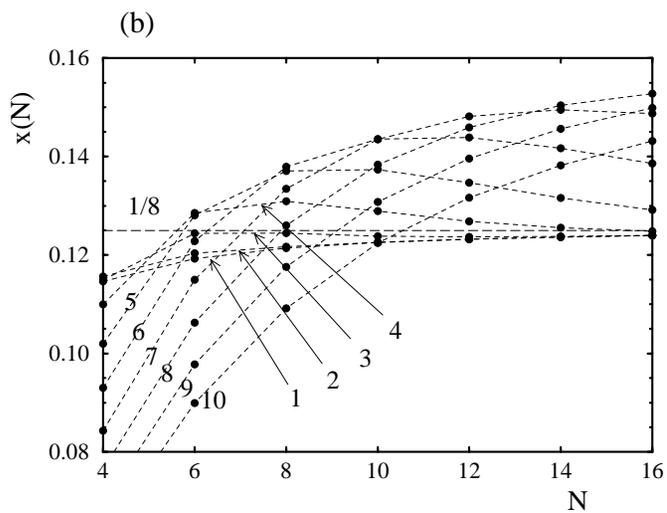
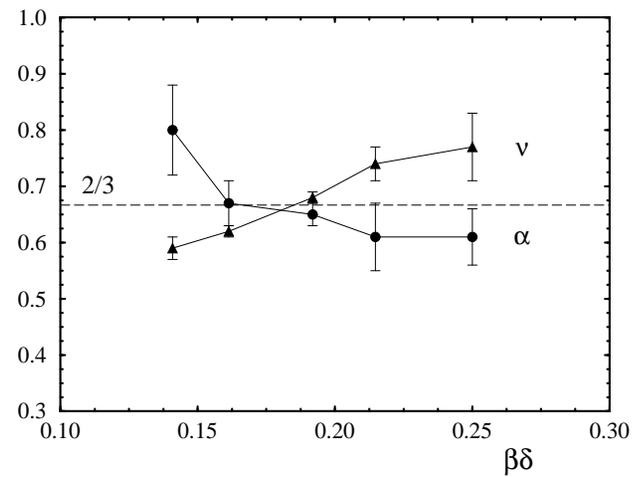
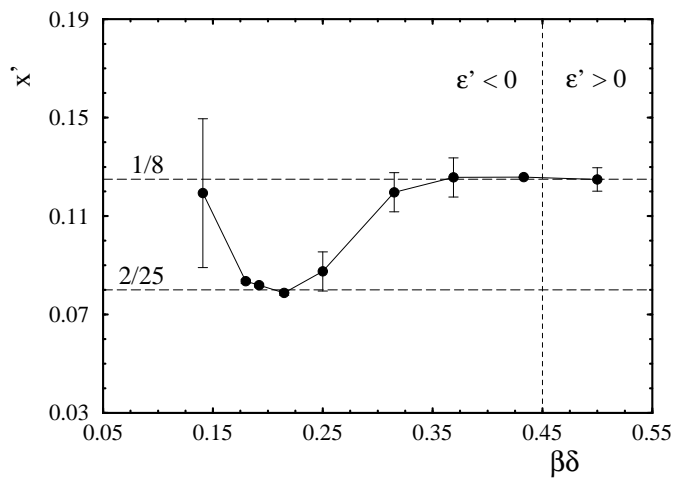
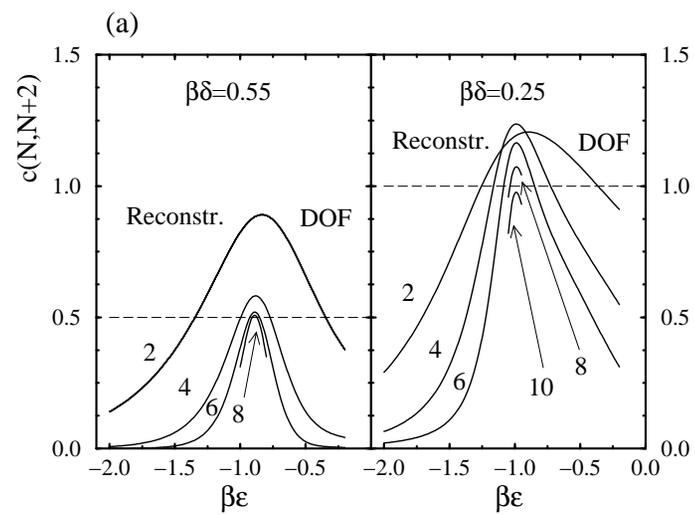
\end{document}